
\input phyzzx.tex
\def\kyotoFIHS{\centerline{\sl Department of Fundamental Sciences}
	\centerline{\sl Faculty of Integrated Human Studies}
          \centerline{\sl Kyoto University, Kyoto 606-01, Japan}}
\def\kyotoFacS{\centerline{\sl Department of Physics, Faculty of Science}
          \centerline{\sl Kyoto University, Kyoto 606-01, Japan}}
\def\NP{Nucl.~Phys.~}
\def\PR{Phys.~Rev.~}
\def\PRL{Phys.~Rev.~Lett.~}
\def\AP{Ann.~Phys.~}
\def\PL{Phys.~Lett.~}
\def\PROG{Prog.~Theor.~Phys.~}
\def\MP{Int.~J.~Mod.~Phys.~}
\def\JMP{J.~Math.~Phys.~}

\catcode`\@=11 
\newtoks\Pubnump   \let\pubnump=\Pubnump
\newtoks\Pubnumpr   \let\pubnumpr=\Pubnumpr
\def\p@bblock{\begingroup \tabskip=\hsize minus \hsize
   \baselineskip=1.5\ht\strutbox \topspace-2\baselineskip
   \halign to\hsize{\strut ##\hfil\tabskip=0pt\crcr
   \the\Pubnum\crcr\the\pubnump\crcr\the\pubnumpr\crcr\the\date\crcr}\endgroup}
\def\author#1{\titlestyle{\fourteencp #1}\nobreak}
\def\abstract{\par\dimen@=\prevdepth \hrule height\z@ \prevdepth=\dimen@
   \vskip 5mm \centerline{\twelverm ABSTRACT}}
\def\title#1{\vskip 1cm \titlestyle{\seventeenbf #1} \vskip\headskip }
\catcode`\@=12 

\VOFFSET = 1.2cm
\HOFFSET = .7cm
\Pubnum{KUCP-0075}
\pubnump{KUNS-1311 HE(TH)94/17}
\pubnumpr{HEP-TH/9412093}
\date{December 1994}
\titlepage
\title{\doublespace Complex-time path-integral formalism for quantum tunneling}
\author{Hideaki Aoyama
\foot{E-mail address: aoyama@phys.h.kyoto-u.ac.jp}}
\kyotoFIHS
\vskip 2mm
\centerline{and}
\author{Toshiyuki Harano
\foot{E-mail address: harano@gauge.scphys.kyoto-u.ac.jp}}
\kyotoFacS
\midinsert\narrower\noindent
\abstract
The complex-time formalism is developed in the framework of the path-integral
formalism,  to be used for analysis of the quantum tunneling phenomena.
We show that subleading complex-time saddle-points do not account for the right
WKB result. Instead, we develop a reduction formula, which enables us to
construct Green functions from simple components of the potential, for which
saddle-point method is applicable. This method leads us to the valid WKB
result, which incorporates imaginary-time instantons and bounces, as well as
the real-time boundary conditions.
\endinsert

\vskip 3cm
\endpage
\normalspace

\chapter{Introduction}

The imaginary-time path-integral method
has been successful for the treatment of
the quantum tunneling phenomena in quantum field theories.
The existence of the solution of the field equation,
instantons and bounces, allows us to apply semi-classical
approximation valid in small coupling.

In recent years, much effort has been made to develop this method further.
One of the driving force was the hope of observing baryon and
lepton number violation process in collision experiments in the TeV range.
This process is associated with the tunneling between topologically different
vacua in the standard electroweak model through the
baryon and lepton number anomaly.
't Hooft\Ref\thooft{G.~'t Hooft \journal \PR &D14 (76) 3432.}
first noticed this possibility, but
at the same time found that the probability is highly suppressed.
Later it was found that the potential barrier
that separated the vacua has a ``pass" (the Sphaleron), at the  height of
few TeV. \Ref\manton{N.~S.~Manton  \journal \PR &D28 (83) 2019;
F.~R.~Klinkhamer and N.~S.~Manton \journal \PR &D30 (84) 2212.}
This implies that even if the tunneling at the bottom of the
potential well is highly suppressed as was estimated by 't Hooft,
it may not be suppressed once initial energy is of order of TeV.
This expectation lead to series of works on developing the original instanton
methods to be applicable to higher energy tunneling.
This development has important implications in quantum theory
in general, including nuclear fission problems
as well as a number of tunneling problems in matter physics.

The original dilute-instanton-gas approximation is valid at the
ground state.  This can be confirmed by comparing the
result in quantum mechanics with that of the WKB calculation.
As the initial energy goes up, instanton result becomes invalid.
In quantum mechanics, it fails in the estimate of the energy splitting of the
first (and higher) excited state in double well.
In quantum field theory, the cross sections violate the unitarity.\Ref\re{
A.~Ringwald \journal \NP &B330 (89) 1;
O.~Espinosa \journal \NP &B343 (90) 310.}
One of the present authors (H.A.) and Kikuchi\Ref\ak{
H.~Aoyama and H.~Kikuchi \journal \PL &B247 (90) 75
\journal \PR &D43 (91) 1999 \journal \MP &A7 (92) 2741.}
found that the interactions between the instantons become
relevant: It gives the right energy splitting in quantum
mechanics and satisfies the unitarity bound for the cross section.

As the energy yet goes up, the distances between instanton and
anti-instanton decreases. Thus the semi-classical configurations
are far from the solution of the equations of motion.
Consequently, the valley methods\Ref\yung{
A.~V.Yung \journal \NP &B191 (81) 47.} were adopted.
The new valley method\Ref\newvalley{
H.~Aoyama and H.~Kikuchi \journal \NP &B369 (92) 219.}
was developed for this purpose in mind
and yields configurations that allows exact conversion of the
smallest eigenvalue to collective coordinate.
Khoze and Ringwald\Ref\kr{
V.V.~Khoze and A.~Ringwald \journal \NP &B355 (91) 351.}
used a series of approximate valley configurations
to estimate the total tunneling cross section.
The new valley method was recently applied to the
bounce problem by one of the authors and Wada.
They solved for the bounce-valley and identified
small bubble configurations valid for
the finite energy decay.\Ref\aw{H.~Aoyama and S.~Wada, {\sl Kyoto University
preprint} KUCP-0069 (1994).}

All these analyses were carried out in the imaginary-time formalism.
Thus basic as well as practical questions related to analytic continuation
remain.
How does the boundary condition matches that of the valley configurations?
In estimating many-point Green functions, how does
the external fields enter (affect) the imaginary time configuration?
[This is crucial for the validity of the
path-deformations in the valley calculation, as well as bounce calculations.]
These are some of the relevant questions.
Some other points are elaborated by Boyanovsky, Willey
and Holman.\Ref\boy{
D.~Boyanovsky, R.~Willey and R.~Holman \journal \NP &B376 (92) 599.}
They also note that in some cases the imaginary-time method
leads us to apparently contradictory results.

In quantum mechanics, the complex-time method has
been studied by various authors.
\REFS\mcl{D.~W.~McLaughlin \journal\JMP &13 (72) 1099.}
\REFS\bender{I.~Bender, D.~Gomez, H.~Rothe and K.~Rothe \journal\NP
&B136 (78) 259.}
\REF\levi{S.~Levit, J.~W.~Negele and Z.~Paltiel \journal\PR &C22 (80) 1979.}
\REF\patra{A.~Patrascioiu \journal\PR &D24 (81) 496.}
\REF\lap{A.~Lapedes and E.~Mottola \journal\NP &B203 (82) 58.}
\REF\weiss{U.~Weiss and W.~Haeffner \journal\PR &D27 (83) 2916.}
\REF\carlitz{R.~D.~Carlitz and D.~A.~Nicole \journal \AP &164 (85) 411.}
\refmark{\boy- \carlitz}
It was argued that it allows semi-classical approximation
for tunneling phenomena and overcomes various problems
associated with the pure-imaginary-time method.\refmark\boy\
In it, one considers the analytic continuation of the
time-integral in Fourier transform of the Feynman kernel.
The existence of the tunneling leads to the
existence of the complex saddle-points in the time-plane.
It was claimed that semi-classical approximation is done by deforming the
time-integral so that it goes through all such saddle-points.\refmark{\carlitz}

This gives one hope that calculations in field theory
may be improved by using the complex-time method.
Indeed, Son and Rubakov
\REFS\khleb{S.~Y.~Khlebnikov, V.~A.~Rubakov and P.~G.~Tinyakov
\journal\NP &B367 (91) 334.}
\REFSCON\son{V.~A.~Rubakov, D.~T.~Son and P.~G.~Tinyakov \journal\PL &287
(92) 342.}
\REFSCON \rubakoc{D.~T.~Son and V.~A.~Rubakov \journal \NP &B424 (94)
55.}
\refsend
adopted a complex-time method for the estimate of the cross section.
They used periodic instanton-anti-instanton solution
in imaginary time, and the initial Minkowski state
overlaps with the instanton as a coherent state,
much like the Aoyama and Goldberg\Ref\ag{H.~Aoyama and H.~Goldberg
\journal \PRL &188B (87) 506.} calculation of the Sphaleron cross section.

Closer look, however, reveals many unanswered questions.\Ref\ahletter{
H.~Aoyama and T.~Harano, {\sl Kyoto University preprint},
KUCP-0072/KUNS-1302 (1994)}
It is known that the complex-time plane is plagued
with singularities and infinite number of saddle-points, which form lattice
structure.
Among saddle-points, there are two kinds, those that are
solutions of the equation of motions (we shall call these
``physical saddle-points" in this paper),
and those that are not (``unphysical saddle-points").
How the path is deformed to avoid the singularities
and to go through only the physical saddle-points is unknown.
One simply assumes that all and only the physical saddle-points contribute.
Even so, the weight of each saddle-point is a riddle.
They are determined so that the result agrees with that of the WKB
approximation.
It is not known how it comes about from the path.

In this paper, we give the solution of these problems,
based on a reduction formula in the number of the
turning points of the path.
In the next section, we give a brief overview
of the saddle-point calculations in complex-time.
There we elaborate some of the points briefly mentioned above.
In section 3, we analyze the orthonormality of the WKB wavefunctions
and construct the Green function from them.
The connection formula for the WKB wavefunction
is discussed in the section 4, where it leads to the
reduction formula for Green function.
Expanding this formula, we obtain a series,
which can be interpreted as a sum over the physical
saddle-points.  The weights and the phases of saddle-points
are determined from this formula.
The last section gives summary and discussions.

\chapter{Saddle-point method}

\FIG\potshape{A potential $V(X)$ with asymptotic regions I and II.}
\FIG\saddles{Positions of saddle-points in complex $T$-plane.}
\def\pri{{^\prime}}

We consider one-dimensional quantum mechanics with action;
$$S = \int dt \left[ {1\over 2} \dot{x}^2 - V(x) \right]_. \eqn\action$$
The potential $V(x)$ is assumed to be smooth enough to
allow WKB approximation in asymptotic regions I $(x \rightarrow -\infty$)
and II $(x \rightarrow \infty$).

The finite-time Green function (Feynman kernel) is defined by the
following in the Heisenberg representation;
$$G(x_i, x_f; T) = \bra{x_f} e^{-iHT} \ket{x_i},
\eqn\tgreen$$
in terms of the hamiltonian of the system, $H$.
The advanced and retarded resolvents are, respectively;
$$\eqalign{
	G^A(x_i, x_f; E) &=
		i \int_{-\infty}^0 dT e^{i(E - i\delta)T} G(x_i, x_f; T)
		= \bra{x_f} {1 \over E-i\delta -H} \ket{x_i}, \crr
	G^R(x_i, x_f; E) &=
		-i \int_0^\infty dT e^{i(E + i\delta)T} G(x_i, x_f; T)
		= \bra{x_f} {1 \over E+i\delta -H} \ket{x_i},}
\eqn\resolvents$$
where real-positive $\delta$ is introduced to guarantee the
convergence of the integrals.
The poles of these Green functions come from the bound states,
and the cut from the continuous spectra.

Let us first look at the saddle-point approximation for the path
integral,
$$G(x_i, x_f; T)=\int_{x(0)=x_i}^{x(T)=x_f} {\cal D} x e^{iS}. \eqno\eq$$
One obtains the following from the saddle-point method;
$$
G(x_i, x_f; T) = \sum_{x_{cl}}\left[{2 \pi \over i}{\dot x_{cl}(0)
\dot x_{cl}(T) \over -\partial^2 S_{cl} / \partial
T^2}\right]^{-1 / 2} e^{iS_{cl}},
\eqn\grnt $$
where $x_{cl}$ is the solution of the classical equation of motion and
$S_{cl}$ is the action evaluated at the classical solution. This
approximation is valid as long as both $x_i$ and  $x_f$ are far from the
turning points where classical velocities vanish.
Substituting \grnt\ into the retarded resolvent in \resolvents,
one arrives at the following expression;
$$ G^R(x_i, x_f; E) = -i \int_0^\infty dT \sum_{x_{cl}} \left[{2\pi \over
i}\ {\dot x_{cl}(0) \dot x_{cl}(T) \over -\partial^2 S_{cl} / \partial
T^2}\right]^{-{1/ 2}}e^{i\left( ET + S_{cl}\right)}. \eqn\green
$$
In the above, we absorbed the infinitesimal imaginary convergence factor
($i\delta$) in $E$.

In calculating \green, one hopes to apply the saddle-point
approximation to the $T$-integral.
The saddle-point condition (the stable-phase condition, to be more
exact) is then,
$$
E=-{\partial S_{cl}\over \partial T} =
{1 \over 2} \dot{x_{cl}}^2 + V(x_{cl}) \ .
\eqn\tcondition
$$
In other words, the solution of the equation of motion $x_{\rm cl}(t)$
has to have the energy $E$.

Let us consider a situation depicted in Fig.1.
Both the initial point $x_i$ and the final point $x_f$
lies in the forbidden region.
There is no real-time solution of the equation of motion.
Thus no semi-classical approximation is possible for
the real-time formalism.

If one assumes
that the expression \grnt\ gives correct analytic continuation
of the Green function in the complex $T$-plane when
complex-$T$ solutions of the equation of motion exist,
one could go to the complex $T$-plane and apply the saddle-point method.
One then arrives at the following expression;
$$
G(x_i, x_f; E) = -i\sum_{T_s} \sum_{x_{cl}} \vert \dot x_{cl}(0)
\dot x_{cl}(T_s)
\vert^{-{1/2}}\ w(T_s)\ p(T_s)\  e^{i W_{cl}(T_s)} ,\eqn\grn
$$
where $T_s$ is given by \tcondition.
Here $W_{cl}(T_s)$ is the WKB phase, $w(T_s)$ is weight and $p(T_s)$ is
phase for the
corresponding saddle-point. The original contour is deformed to pass
the saddle-points by a series of steepest descent paths
from the origin $T=0$ to
the positive infinity on the real axes (Re $T=\infty$, Im $T=0$).
The weight of a saddle-point, $w(x_{cl})$,
is determined by how the contour crosses
the saddle-point: If the contour does not cross the saddle-point, the
corresponding weight is zero. If the contour crosses it along
the steepest descent direction, the weight is $1$.
In case the contour reaches a sub-dominant saddle-point from the
direction orthogonal to the steepest descent
and leaves it along the steepest descent direction, the weight is $1/2$
(as in the calculation of the false vacuum decay).
The phase of a saddle-point, $p(T_s)$ comes from the square root in
Eq.\green.

Carlitz and Nicole applied this method to
linear potential, quadratic well and quadratic barrier, all of which
are exactly solvable.
Using the exact expression for the Green function \tgreen,
they found that the saddle-points explained above, namely
the ones associated with the complex-time solution, indeed
lead to saddle-points in the $T$-plain
[we shall call these ``physical saddle-points"].
They, however found more: there were saddle-points
that are {\sl not} associated with any solutions of
equations of motion (unphysical ones), and also singularities due to the
periodicities.
Deforming the $T$-integration path to avoid singularities,
they found that the path indeed goes through only the physical saddle-points.
As a result, the weights and phases in \grn\ are determined,
which lead to the correct WKB result.

In the case of a double-well potential, the deformation of the
integration contour is not specified,
which is understandable in view of the fact that
the analytic structure of \tgreen\ in complex $T$-plain
is fairy complicated.
The claim made in the literatures is simply that
the contour passes all saddle-points that correspond to
classical trajectories with weights that are extracted
from the above calculation.

The problem lies in the fact that we do not  know the contour which passes
all physical saddle-points that are distributed in complex-time plane.
Because of this, we cannot determine the weights of the saddle-points.
In the case where $x_i$ and $x_f$
are in the forbidden region on the right side of the double-well
potential as depicted in Fig.\potshape, saddle-points in complex
$T$-plane are given by
$$
\eqalign{T_s =\pm T(b_2,x_i) \pm T(b_2,x_f) + l T(a_1,b_1) +& mT(b_1,a_2) +
nT(a_2,b_2)  \crr
 & l,m,n = 0,\pm 1, \pm 2
,...\ , }\eqno\eq
$$
where
$$
T(x,y) = 2 \int_{x}^{y} {dx\pri  \over  \sqrt {2 \left( E-V(x\pri) \right)}}.
\eqno\eq
$$
Notice that $T(x,y)$ is pure imaginary if $(x, y)$ is in a forbidden
region. These saddle-points $T_s$ are
shown as solid circles in Fig.\saddles.
The unphysical saddle-points are also shown as open circles in
the same figure. We know of no contour
which passes only the physical ones and avoids the unphysical ones.
Thus so far the complex-time method has not been worked out for
the double-well and more complicated potentials.

\chapter{Green function and the WKB wavefunctions}

We shall examine the Green functions by using the
complete orthonormal set of the eigenfunctions $\{ \ket{\psi_n}\}$ of $H$.
For example, the retarded Green function in \resolvents\ is written as
$$G^R(x_i, x_f; T) = \sum_n {\braket{x_f}{\psi_n} \braket{\psi_n}{x_i}
	 \over E + i\delta -\lambda_n}, \eqn\advg$$
where the sum over the complete set is actually made of the integration
over the continuum spectrum and sum over the discrete bound states.

\def\bl{_\lambda}

In order to study of the Green functions in the asymptotic region,
we use the WKB approximation for the wavefunctions
$\psi(x) = \braket{x}{\psi}$.
For a state with the eigenvalue $\lambda$ in the continuum spectrum,
the second-order WKB approximation yields the following;
$$\psi_\lambda (x) = { 1 \over \sqrt{2\pi p(x)} } \times \cases{
	\left(A\bl e^{i \int_x^{x_1} p(x\pri) dx\pri}
	+ B\bl e^{-i \int_x^{x_1} p(x\pri) dx\pri}\right)_, &for $x\in$ I, \crr
	\left(C\bl e^{-i \int_{x_2}^x p(x\pri) dx\pri}
	+ D\bl e^{i \int_{x_2}^x p(x\pri) dx\pri}\right)_,
		&for $x \in$ II, \cr}\eqn\wavef$$
where $p(x) = \sqrt{2(\lambda-V(x))}$.
We choose the limits $x_1$ and $x_2$ of the integrations
to be the nearest turning points for definiteness.
Among the coefficients $A$, $B$, $C$, and $D$, only two
are independent. Their inter-relation is linear due to the
superposition principle and can be written
as follows;\Ref\ak{H. Aoyama and M. Kobayashi \journal \PROG &64 (80) 1045.}
$$\pmatrix{A\bl \crr B\bl} = S(\lambda) \pmatrix{C\bl \crr D\bl}_. \eqno\eq$$
The 2$\times$2 matrix $S(\lambda)$ is determined by the
shape of the potential $V(x)$ in the intermediate region between
$x_1$ and $x_2$.
The flux conservation law is satisfied for the Schr\"odinger equation;
$$ {d\over d x} j = 0, \quad
	j = {i \over 2} \left({d \psi^\ast \over d x } \psi
	    - \psi^\ast {d \psi \over d x }\right)_.  \eqn\flux$$
Substituting the WKB expression \wavef\ into the above, we find that
$$ j = {1\over 2 \pi} (|A|^2 - |B|^2) =
	{1\over 2 \pi} (|C|^2 - |D|^2). \eqn\abcd$$
This results in the following relation between the matrix elements of
$S(\lambda)$;
\def\se#1{S_{#1}}
\def\sen#1{|S_{#1}|^2}
$$\eqalign{\sen{11} - \sen{21} &= 1, \quad \sen{12} - \sen{22} = 1, \crr
	\se{11}^\ast \se{12} &- \se{21}^\ast \se{22} = 0.} \eqn\sesen$$
As a result, the matrix $S(\lambda)$ is parametrized by three real
functions $\alpha(\lambda), \beta(\lambda),$ and $\rho(\lambda)$ as follows;
$$S(\lambda) = \pmatrix{ e^{i\alpha} \cosh\rho & e^{i\beta} \sinh\rho \crr
e^{-i\beta} \sinh\rho & e^{-i\alpha} \cosh\rho }. \eqn\sabrho$$

For a given eigenvalue $\lambda$, there are two independent eigenfunctions.
In order to construct them, we look at the inner-product of the
eigenfunction $\psi_\lambda (x)$ with coefficients
($A_\lambda, B_\lambda, C_\lambda, D_\lambda$)
and another eigenfunction  $\tilde \psi_{\lambda\pri} (x)$ with
($\tilde A_{\lambda\pri}, \tilde B_{\lambda\pri}, \tilde C_{\lambda\pri},
\tilde D_{\lambda\pri}$).
Integration in the asymptotic regions determines the coefficient
of the delta function $\delta(\lambda - \lambda\pri)$ completely, as
the delta function can come only from the infinite integrations.
[In order to calculate the finite terms we need to solve the
Schr\"odinger equation completely. But this is not necessary for the
current purpose.]
In fact, integration in the region II yields,
$$\eqalign{
\int_{x_2}^\infty d x \ &\psi^\ast_\lambda (x)\tilde \psi_{\lambda\pri} (x)
	= {1 \over 2\pi} \int_{x_2}^\infty
		{d x \over \sqrt{p(x) p\pri (x)}}\crr
	\times\Bigg( &C_\lambda^\ast \tilde C_{\lambda\pri}
		e^{i \int_{x_2}^x ( p(x\pri) -  p\pri (x\pri) ) d x\pri}
	+ D_\lambda^\ast \tilde D_{\lambda\pri}
		e^{-i \int_{x_2}^x ( p(x\pri) - p\pri (x\pri) ) d x\pri}\crr
	&+ C_\lambda^\ast \tilde D_{\lambda\pri}
		e^{i \int_{x_2}^x ( p(x\pri) + p\pri (x\pri) ) d x\pri}
	+ D_\lambda^\ast \tilde C_{\lambda\pri}
		e^{-i \int_{x_2}^x ( p(x\pri) + p\pri (x\pri) ) d x\pri}
	\Bigg)
}
\eqn\ortho$$
As we are interested in the singularity of the above
for $\lambda = \lambda\pri$, we expand the exponent
using the following for $\lambda \sim \lambda\pri$,
$$p(x\pri) -  p\pri (x\pri) \simeq { (\lambda - \lambda\pri) \over p(x\pri)}.
\eqn\simp$$
By using a new coordinate $y$ defined by
$ d y = d x / p(x)$, we find the contribution from
the first two terms to the delta function;
$$\eqalign{
\int_{x_2}^\infty d x \ \psi^\ast_\lambda (x)\tilde \psi_{\lambda\pri} (x)
	&= {1 \over 2\pi} \int_{y_2}^\infty d y
	\left( C_\lambda^\ast \tilde C_{\lambda}
        e^{i (\lambda-\lambda\pri) (y-y_2)} +
	D_\lambda^\ast \tilde D_{\lambda}
		e^{-i (\lambda-\lambda\pri) (y-y_2)} \right)+ ... \crr
	&= {1 \over 2} \left( C_\lambda^\ast \tilde C_{\lambda}+
	D_\lambda^\ast \tilde D_{\lambda} \right)
	\delta(\lambda - \lambda\pri) + ... }
\eqn\almost$$
where we have neglected all the finite terms.
Doing the similar calculation for the region I, we find
$$\int_{-\infty}^\infty d x \psi^\ast_\lambda (x)\tilde \psi_{\lambda\pri} (x)
 = {1 \over 2} \left(
	A_\lambda^\ast \tilde A_{\lambda}+
	B_\lambda^\ast \tilde B_{\lambda}+
	C_\lambda^\ast \tilde C_{\lambda}+
	D_\lambda^\ast \tilde D_{\lambda} \right)
	\delta(\lambda - \lambda\pri). \eqn\combined$$
Using the above result, we choose our orthonormal eigenfunctions
for a given $\lambda$ as the following;
$$\eqalign{
	(A_\lambda, B_\lambda, C_\lambda, D_\lambda)^{(1)} &=  \left(
	{e^{i\alpha} \over \cosh \rho}, \ 0, \ 1, \
	-e^{i(\alpha - \beta)}\tanh\rho
	\right), \crr
	(A_\lambda, B_\lambda, C_\lambda, D_\lambda)^{(2)} &=  \left(
	e^{i(\alpha + \beta)}\tanh\rho, \ 1, \ 0, \
	{e^{i\alpha} \over \cosh \rho}
	\right).}
\eqn\choice$$
[The above corresponds to two incoming states.
There are of course other choices, such as stationary states,
but all of those reads to the same result for the Feynman kernel.]

\FIG\lambdaplane{The complex-plane of the eigenvalue $\lambda$.
The cut on the real-axis is the continuous spectra and
the poles are the bound states.
The integration contour $C$ is for the continuous part of the Green
function. The discreet part can be represented as integration
around poles.}

Let us first look at the  case when both $x_i$ and $x_f$ are in
an allowed region in II.
{}From \choice, we find that
$$\eqalign{\sum_{i=1,2}\psi_\lambda^{(i)}(x_f)&\psi_\lambda^{(i)\ast}(x_i)
={1\over 2\pi \sqrt{p(x_i)p(x_f)}}\crr
\times &\left[
e^{i \int_{x_i}^{x_f} p(x\pri) dx\pri}
 - e^{i \left(
\int_{x_2}^{x_i} p(x\pri) dx\pri +\int_{x_2}^{x_f} p(x\pri) dx\pri \right)}
e^{i(\alpha - \beta)} \tanh \rho + ({\rm c.c.}) \right]_. \crr}  \eqn\xxintwo$$
The complex conjugate part can be understood as the
same function below the cut on the real-axis of the complex
$\lambda$-plane (see Fig.\lambdaplane).
This is guaranteed by the fact that
as $\lambda$ moves below the cut, the role of the
coefficients $A$ and $B$ is exchanged and likewise for $C$ and $D$,
and as a result, the phases $\alpha$ and $\beta$ change their signs,
while $\rho$ does not.
Therefore the sum over the continuous spectra in the Feynman kernel \advg\
is written as the following $\lambda$-integration;
$$\eqalign{G^R&(x_i, x_f;E) =
{1\over 2\pi}  \int_C d\lambda \
{1 \over E + i\delta - \lambda}
{1 \over \sqrt{p(x_i)p(x_f)}}
\crr
\times &
\cases{
e^{i \int_{x_2}^{x_i} p(x\pri) dx\pri}
\left[e^{-i \int_{x_2}^{x_f} p(x\pri) dx\pri}
- e^{i \int_{x_2}^{x_f} p(x\pri) dx\pri}
e^{i(\alpha - \beta)} \tanh \rho\right] &for $x_i > x_f$ \crr
e^{i \int_{x_2}^{x_f} p(x\pri) dx\pri}
\left[e^{-i \int_{x_2}^{x_i} p(x\pri) dx\pri}
- e^{i \int_{x_2}^{x_i} p(x\pri) dx\pri}
e^{i(\alpha - \beta)} \tanh \rho\right] &for $x_f > x_i$  . \cr}}
  \eqn\gxxintwo$$
The integration contour $C$ in the complex $\lambda$-plane
is as in Fig.\lambdaplane.
In \gxxintwo, we have chosen the integrand so that it
converges for $|\lambda| \rightarrow \infty$,
which we need later.
As for the discreet spectrum, it is determined by the
absence of the diverging behaviour of the wavefunction,
which translates into the condition that for $C=0$, $A=0$.
{}From \sabrho, we thus find that $\sinh \rho =0$ when $\lambda$
is equal to a discreet eigenvalue of $H$.
This allows us to write the sum over the discreet spectra
as pole integrations of the second term in \gxxintwo\
depicted in Fig.\lambdaplane.
Connecting all the contours and closing it at the
infinity, we find that the whole contour enclose the
pole at $\lambda=E + i \delta$.  Thus we end up with the expression
$$\eqalign{G^R(x_i, x_f;E) &=
 - {i \over \sqrt{p(x_i)p(x_f)}} \crr
\times &
\cases{
e^{i \int_{x_2}^{x_i} p(x\pri) dx\pri}\left(e^{-i \int_{x_2}^{x_f}
p(x\pri) dx\pri}
+ iR e^{i\int_{x_2}^{x_f} p(x\pri) dx\pri} \right) &for $x_i > x_f$, \crr
e^{i \int_{x_2}^{x_i} p(x\pri) dx\pri}\left(e^{-i \int_{x_2}^{x_f}
p(x\pri) dx\pri}
+ iR e^{i\int_{x_2}^{x_f} p(x\pri) dx\pri} \right) &for $x_f > x_i$. \crr}}
\eqn\ggxxintwo$$
where we defined the ``reflection coefficient'' $R$ by the following,
$$ R = i {S_{21} \over S_{22}}
= i e^{i(\alpha - \beta)} \tanh \rho . \eqn\rdef$$
We note that this expression \ggxxintwo\ is valid for
a general value of $E$ with proper analytic continuation of
the coefficients.
Therefore, if any end-points are in a forbidden region,
simple analytic continuation of \ggxxintwo\ is appropriate.
We shall use this result in the next section.

\chapter{Reduction formula and its expansion}

\FIG\one{A potential with $n$ wells we consider.
The turning points are denoted by $a_i$ and $b_i$,  for left
and right of the $i$-th well, respectively ($i = 1 \sim n$).}
\FIG\two{Diagrammatic representation of the expansion of $iR_n$.
The first term corresponds to the direct reflection at the turning
point $b_n$. Each path in the rest corresponds to the group of paths that
have a number of oscillation in the $n$-th well.
The information of the region left of $a_n$ is contained in $-i\tilde R_n$.}
\FIG\five{Diagrammatic representation of the expansion of
$-i\tilde R_n$. The each path corresponds to a group of paths
with different number of oscillation in the forbidden region
$(b_{n-1}, a_n)$.}
\FIG\eight{Classical paths traversing in the $n$-th well
and tunneling through the $(n-1)$-th barrier.
These are again expanded into groups of paths with different
oscillations in either the allowed or forbidden regions.}
\FIG\ten{A metastable system and a scattering process against a
potential barrier.}

In this section, we derive a formula for evaluating the Green
function in a system with an arbitrary potential to which we can apply
the WKB approximation.
We consider a general potential which has arbitrary number of
wells, and show that the Green function is given by
summing up all contributions of classical paths.

Let us consider the case where $x_i$ and  $x_f$ lie in the forbidden
region on the same sides of the wells.
The potential which has $n$ wells is depicted
in Fig.\one.
By the analytical continuation of \ggxxintwo\ in $E$,
we find the retarded resolvent in the following form;
$$
 G^R(x_i, x_f; E) = -\vert p(x_i) p(x_f) \vert^{-{1/2}}
 e^{- \Delta_i} \left( e^{ \Delta_f}+iR_n e^{-
 \Delta_f}\right)_,
 \eqn\nwell
$$
where
$$\Delta_{i,f} \equiv \int_{b_n}^{x_{i,f}} dx |p(x)|. \eqno\eq$$
In order to specify the fact that this expression is for
$n$ wells, we attach the subscript $(n)$ to the coefficients
hereafter.

Due to the existence of the
intermediate region $(b_{n-1},a_n)$, the matrix $S^{(n)}$,
which connects the regions I and II, can be written in terms of the
matrices $S^{(n-1)}$ that connects regions I and $(b_{n-1},a_n)$ and
$\tilde{S}$ for $(b_{n-1},a_n)$ and II. If we apply the linear WKB
connection formula for the latter region,
which we can obtain from the saddle-point method for a linear
potential,  we obtain the following;
$$
 \eqalignno{S^{(n)}(E) &= S^{(n-1)}(E)
       \pmatrix{ e^{- \Delta_{n-1}} & 0 \cr
                              0 & e^{ \Delta_{n-1}} \cr}
       \pmatrix{ {i \over 2} & -{i \over 2} \cr
                         1   &   1 \cr}\cr
& \hskip 2cm \times
       \pmatrix{e^{-i \left( W_{n}+{ \pi \over 2} \right) } & 0 \cr
                0 & e^{i \left( W_{n}+{ \pi \over 2} \right )}  \cr}
       \pmatrix{{1 \over 2} &  i \cr
                {1 \over 2} & -i \cr} \crr
 &= S^{(n-1)}(E)
       \pmatrix{
      {1 \over 2} e^{- \Delta_{n-1}}\cos W_{n}
  &   e^{- \Delta_{n-1}} \sin W_{n} \cr
     -e^{ \Delta_{n-1}} \sin W_{n}
  &   2 e^{ \Delta_{n-1}}\cos W_{n}\cr}
 &\eq\cr
 }
$$
where
$W_{n}$\  and $\Delta_{n-1}$ \ are defined by,
$$
 W_{n} = \int_{a_{n}}^{b_{n}}dx |p(x)| ,
 \eqn\wn
$$
$$
 \Delta_{n-1} = \int_{b_{n-1}}^{a_{n}}dx |p(x)| .
          \eqno\eq
$$
Therefore, we find the relations;
$$
 S_{21}^{(n)} =
  {1 \over 2 }S_{21}^{(n-1)} e^{- \Delta_{n-1}} \cos W_{n}
 -S_{22}^{(n-1)} e^{ \Delta_{n-1}} \sin W_{n}
 \eqn\cn
$$
$$
 S_{22}^{(n)} =
   S_{21}^{(n-1)} e^{- \Delta_{n-1}} \sin W_{n}
 + 2 S_{22}^{(n-1)} e^{ \Delta_{n-1}} \cos W_{n} \ .
 \eqn\dn
$$
{}From \rdef, \cn\ and  \dn, we find that $R_{n}$ can be written as the
following;
$$
 R_{n}  = {i \over 2}\
{- \sin W_{n}-{i \over 2} e^{-2 \Delta_{n-1}} R_{n-1} \cos W_{n}
 \over
 \cos W_{n} -{i \over 2}e^{-2 \Delta_{n-1}} R_{n-1} \sin W_{n} }
\eqno\eq$$
We find it most convenient to rewrite this to the following two
expressions;
$$  R_{n} = {1 \over 2}\ {1-\tilde{R}_{n} e^{2iW_{n}}
                               \over 1+\tilde{R}_{n} e^{2iW_{n}}},
\eqn\gnplusone$$
$$  \tilde{R}_{n}={
 {1-{1 \over 2}R_{n-1} e^{-2 \Delta_{n-1}}}
 \over
 {1+{1 \over 2}R_{n-1} e^{-2 \Delta_{n-1}}}
 }.  \eqn\fn$$
Defined in this manner, the function $R_{n-1}$ corresponds to the
reflection amplitude at the turning point $b_{n-1}$.

In order to see the correspondence between these expressions
and the saddle-point method, let us expand
\gnplusone\ in an infinite series as follows;
$$
iR_{n} =  {i \over 2} + \left( -i\tilde{R}_{n} \right)\ e^{2iW_{n}} +
(-i\tilde{R}_{n})^2 (-i)\ e^{4iW_{n}} + ...  \quad .
\eqn\gnrec
$$
The convergence of this series is guaranteed by the implicit factor
$i \delta$.
The retarded resolvent is then,
$$
\eqalign{
&G^R(x_i, x_f; E) = -|p(x_i) p(x_f)|^{-{1/2}}   \crr
 & \times \left[ e^{-\left( \Delta_i -\Delta_f
\right)} + \left\{{i \over 2}+\left( -i\tilde{R}_{n} \right)\ e^{2iW_{n}} +
\left( -i\tilde{R}_{n} \right)^2(-i)\ e^{4iW_{n}}
+ ... \right\} e^{-(\Delta_i +\Delta_f)} \right].}
\eqn\grexpand
$$
This is the expression to be compared to the saddle-point expression \green.
If we choose $T$ to be negative imaginary, $-i\tau$,
the factor in the exponent is,
$$i(ET+S_{cl}) = E\tau - S_E = -\int d\tau \left( {dx\over
d\tau}\right)^2 = -\int dx |p(x)|\eqno\eq$$
Therefore, the two factors $\Delta_{i,f}$ is equal to the above
quantity for the path from $x_{i,f}$ to the turning point $b_n$.
The first term in \grexpand\ corresponds
the contribution of the pure-imaginary-time path that starts from
$x_i$ and reaches $x_f$ directly.
The factor $e^{-(\Delta_i +\Delta_f)}$ in the rest of the terms
is for the path from $x_i$ to the turning point $b_n$ and then from
$b_n$ to $x_f$.
The expansion of $iR_n$ is understood as contributions of
the paths that oscillate in the allowed region $(a_n, b_n)$.
Various factors have unique interpretations as factors coming
from the turning points.  This is most conveniently depicted in Fig.\two.
Similarly to \gnrec, the expression \fn\ is expanded as the following,
$$
-i\tilde{R}_{n} = -i + (iR_{n-1})\ e^{-2 \Delta_{n-1}}
+ (iR_{n-1})^2 ({i \over 2})\ e^{-4 \Delta_{n-1}} + ...   \quad  .
\eqn\fnrec
$$
The corresponding diagrams are illustrated in Fig.\five.

We have so far derived the expressions that results from the
WKB approximation for the $n$-th well.
This procedure can be applied recursively for the rest of the
wells.  At the end, we are left with $R_1$, which is given by
$$iR_1 =  {1  \over  2} \tan W_1
       = {i \over 2} + (-i)\ e^{-2iW_1} + (-i)^3\ e^{-4iW_1}
+ ...   \eqn\rtildeexpand
$$
where $W_1$ is defined by \wn\ by replacing $n$ with $1$. Therefore we
find that $R_1$ is expressed as a sum of the contribution of the
classical paths which evolve in the first well.

Combining \grexpand, \fnrec, and \rtildeexpand,
we find all complex-time paths are included in the
expression of the resolvent $G^R$.
Therefore, the resulting Green function in a $n$-well potential is given by
$$
G^R(x_i, x_f; E) = - |p(x_i) p(x_f)|^{-1/2} \sum_{x_{cl}}\
f(x_{cl}) e^{i W(x_i,x_f; E)} .\eqn\grn
$$
where $f(x_{cl})$ is determined by the number of reflections and
transmissions which the classical path contains, and $W(x_i,x_f; E)$ is
determined by what wells the path crosses and what barriers it tunnels
through. The rules for calculating $f(x_{cl})$ is the following: When a
path has a reflection in the allowed region, it obtains $-i$. In the
case of a reflection in the forbidden region, it gets $i \over 2$.

Let us next examine the case
when $x_i$, $x_f$ are on opposite sides of the wells.
Applying the analysis similar to the one in the section 3,
we find that the Green function is expressed as follows;
$$
 G^R(x_i, x_f; E) = -\vert p(x_i) p(x_f) \vert^{-{1/2}}
T_n e^{- \Delta_i - \Delta_f}
 \eqno\eq
$$
where $T_n$ is the (analytically-continued) transmission amplitude
$$
 T_n = {1 \over S_{22}^{(n)}} \ .
 \eqno\eq
$$
Just as the previous case, we can express $T_n$ in terms of $T_{n-1}$
as in the following;
$$
 \eqalignno{ T_{n}  &= {1 \over {S_{21}^{(n-1)}} e^{- \Delta_{n-1}}
\sin W_n + 2 S_{22}^{(n-1)}  e^{ \Delta_{n-1}} \cos W_n} \crr
 &= {  T_{n-1} e^{- \Delta_{n-1} + i W_n}
     \over
       1 + {1 \over 2} R_{n-1} e^{-2 \Delta_{n-1}} + e^{2iW_n}\left(1- {1 \over
 2}R_{n-1} e^{-2 \Delta_{n-1}}\right)}  \crr
 &= { e^{- \Delta_{n-1}} \over 1 + {1 \over 2} R_{n-1} e^{-2 \Delta_{n-1}}}
   \ { e^{i W_n}  \over 1 + \tilde{R}_{n} e^{2i W_n}}
   \ T_{n-1}.
 &\eqname{\tn}\cr}
$$
In the above, we have used \rdef , \dn\   and \fn.
Let us show that $T_{n}$ consists of all contributions of classical
paths. The second factor in \tn\ is expanded in the following way,
$$
{e^{iW_n} \over  1+\tilde{R}_n e^{2iW_n}}=e^{iW_n} +
(-i)(-i\tilde{R}_n)\ e^{3iW_n} + (-i)^2(-i\tilde{R}_n)^2\ e^{5iW_n}+...
\quad ,
\eqno\eq
$$
which corresponds to the diagrams in Fig.\eight\ (a). The first factor
in \tn\ is expanded as,
$$
  {  e^{- \Delta_{n-1}} \over 1+{1 \over 2}R_{n-1}\ e^{-2
\Delta_{n-1}}} =
e^{- \Delta_{n-1}} + ({i \over 2})(iR_{n-1})\ e^{-3 \Delta_{n-1}} +  ({i \over
2})^2(iR_{n-1})^2\ e^{-5 \Delta_{n-1}} + ...  \quad,
\eqno\eq
$$
which is illustrated in Fig.\eight\ (b).
As the previous case, repeating this procedure, we reach $T_1$, which
is given by
$$
\eqalignno{T_1 &={1 \over 2 \cos W_1}  \crr
               &=e^{iW_1} + (-i)^2\ e^{3iW_1} + (-i)^4\ e^{5iW_1}+... \quad .
&\eq\cr}
$$
Therefore we conclude that $T_n$ consists of all contributions of
classical paths.

The above analysis can be applied to other cases with various
locations of $x_i$ and $x_f$.
Thus we verify that thus constructed resolvent can be always
interpreted as sum over the complex-time classical paths.

This formalism is valid for a metastable system and scattering process
(Fig.\ten) as well as for a stationary system. It is apparent from
this derivation that this complex-time method reproduces the result
of the WKB approximation with the linear connection formula.
The potential we consider is given in Fig.\ten (a).
The turning points are labeled in the figure by $a_1$, $b_1$, and $a_2$.
First we consider the case when $x_i$ is in the well and $x_f$ is
outside of the barrier.
The analysis similar to the above yields the Green function as in the
following;
$$
\eqalign{
G^R(x_i, x_f; E) = -i |p(x_i) p(x_f)|^{-1/2}
 { \left( e^{-iW_i} -ie^{iW_i} \right)\ e^{iW_1- \Delta_1 +iW_f}
\over (1+\tilde{R} e^{2iW_1})(1+{1 \over 4} e^{-2 \Delta_1})},}
\eqn\metaone
$$
where $\tilde{R}$ is given by
$$
\eqalign{-i\tilde{R}
 &= -i\ {1-{1 \over 4}e^{-2\Delta_1} \over 1+{1 \over
4}e^{-2\Delta_1}} \crr
 &= -i + {i \over 2} e^{-2\Delta_1} +({i \over 2})^3 e^{-4\Delta_1}
+... \ .}\eqn\metatwo
$$
{}From \metaone\ and \metatwo, we find again that the resolvent is
equal to the sum over the physical saddle-points.
We find that the poles of the Green function
are determined by the following;
$$
1+ e^{2iW_1} + {1 \over 4} e^{-2\Delta_1} \left( 1-e^{2iW_1} \right)=0.
\eqno\eq
$$
Let us solve this equation iteratively. Then we obtain
$$
W_1(E) = \left( n+{1 \over 2} \right) \pi -{i \over 4} e^{-2\Delta_1}.
\eqn\metathree
$$
{}From  \metathree, we find an imaginary part of the energy eigenvalue
$$
{\rm Im}\ E_n =  - {i \over 2T(E_n)} e^{-2\Delta_1},
\eqn\imeexpress
$$
where $T(E_n)$ is the period of the classical path between the
turning points $a_1$, $b_1$
$$
T(E) = 2 \int_{a_1}^{b_1} {dx  \over  \sqrt {2 \left( E-V(x) \right)}}.
\eqno\eq
$$
We find the decay rate of this metastable system.
We note that the factor $1/2$ in  \imeexpress\ comes from the
weight $1/2^n$ of the saddle-points with $n$ reflections in the forbidden
region.

Let us apply this method to a scattering process against a potential
barrier in Fig.\ten. When $x_i$ and $x_f$ are separated by the
barrier, the Green function is given by
$$
\eqalignno{G^R(x_i, x_f; E)
 &=-i |p(x_i) p(x_f)|^{-1/2} \tilde{T}e^{i(W_i+W_f)}  \crr
&= -i |p(x_i) p(x_f)|^{-1/2}
\sum^{\infty}_{k=1} e^{i(W_i+W_f)-(2k-1)\Delta_1} \left({i \over
2}\right)^{2(k-1)}_,
&\eqname{\transmission}\cr}
$$
where we have used the WKB expression for the transmission
coefficient $\tilde{T}$;
$$ \tilde{T} =\  {e^{-\Delta_1} \over 1+{1 \over 4}e^{-2\Delta_1}} \eqno\eq$$
When $x_i$ and $x_f$ are on the same side, the Green function is given
by
$$
\eqalignno{&G^R(x_i, x_f; E)
= -i |p(x_i) p(x_f)|^{-1/2}
e^{iW_i}\left[e^{-iW_f}-i \tilde{R} e^{iW_f}\right] \crr
&= -i |p(x_i) p(x_f)|^{-1/2}
e^{iW_i}\left[e^{-iW_f}+{i \over 2}e^{iW_f}+\sum^{\infty}_{k=1}
e^{iW_f-2k \Delta_1} (-i)^{2k-1}\right]_,
&\eqname{\reflection}\cr}
$$
where we have used the expression for the reflection amplitude,
$$
-i \tilde{R}=-i \ {1-{1 \over 4}e^{-2\Delta_1} \over 1+{1
\over 4}e^{-2\Delta_1}}.
\eqno\eq
$$
The coefficients $\tilde{T}$ and $\tilde{R}$ satisfy the unitarity condition
$$
|\tilde{T}|^2+|\tilde{R}|^2=1. \eqn\trunitary
$$
We again confirm the validity of the sum over the physical
saddle-points.
The weight $1/2^n$ is crucial for \trunitary.

\chapter{Discussion}

In the paper, we have given the re-formulation of the complex-time method.
Using the connection formula for the wavefunctions, we constructed the
reduction formula in the number of the turning points
for the Green function.
This yields series expansions, which can be understood as sum over
in the classical complex-time trajectory.
This is understood as a sum over the
physical saddle-points with specific weights and phases in the path-integral
method.
This shows the validity of the method proposed before, in the
context of the path-integral method.
We confirmed that this method yields results identical with
that of the WKB approximation.
Thus our construction gives solid basis for the starting point
of the complex-time method.

What happens if one tries to extend this method to field theories?
Not much is known.  Rubakov et.al. basically assumes that
the behavior of the Green functions in asymptotic region
($T \rightarrow \infty$) does not depend on the
imaginary part of $T$ and calculate with non-zero Im $T$
so that the saddle-point approximation can be applied.
However, even for Re $T \rightarrow \infty$,
the Green function {\sl does} depend on the Im $T$,
as we are allowed to have more configurations along the
imaginary-time axis as Im $T$ increases.
Thus rather careful consideration would be required.
The complex-time method instead relies on the $T$ integration
as we have seen in this method.
Thus somewhat different analysis may be required.
One of them could be the combination of the
valley method and the current complex-time method.
The valley method can used to identify the
imaginary-time tunneling paths, which are
converted to collective coordinates.
Thus, while the incoming particle are expressed in
real-time expressions, the tunneling part
may be obtained as the imaginary part of the
complex-time development along the valley trajectory.
There are rather interesting possibilities along this line,
which should be further investigated.
\endpage

\refout\endpage\figout
\endpage

\input epsf

{}~~
\vfill
\centerline{
\epsfxsize=8.3cm
\epsfbox{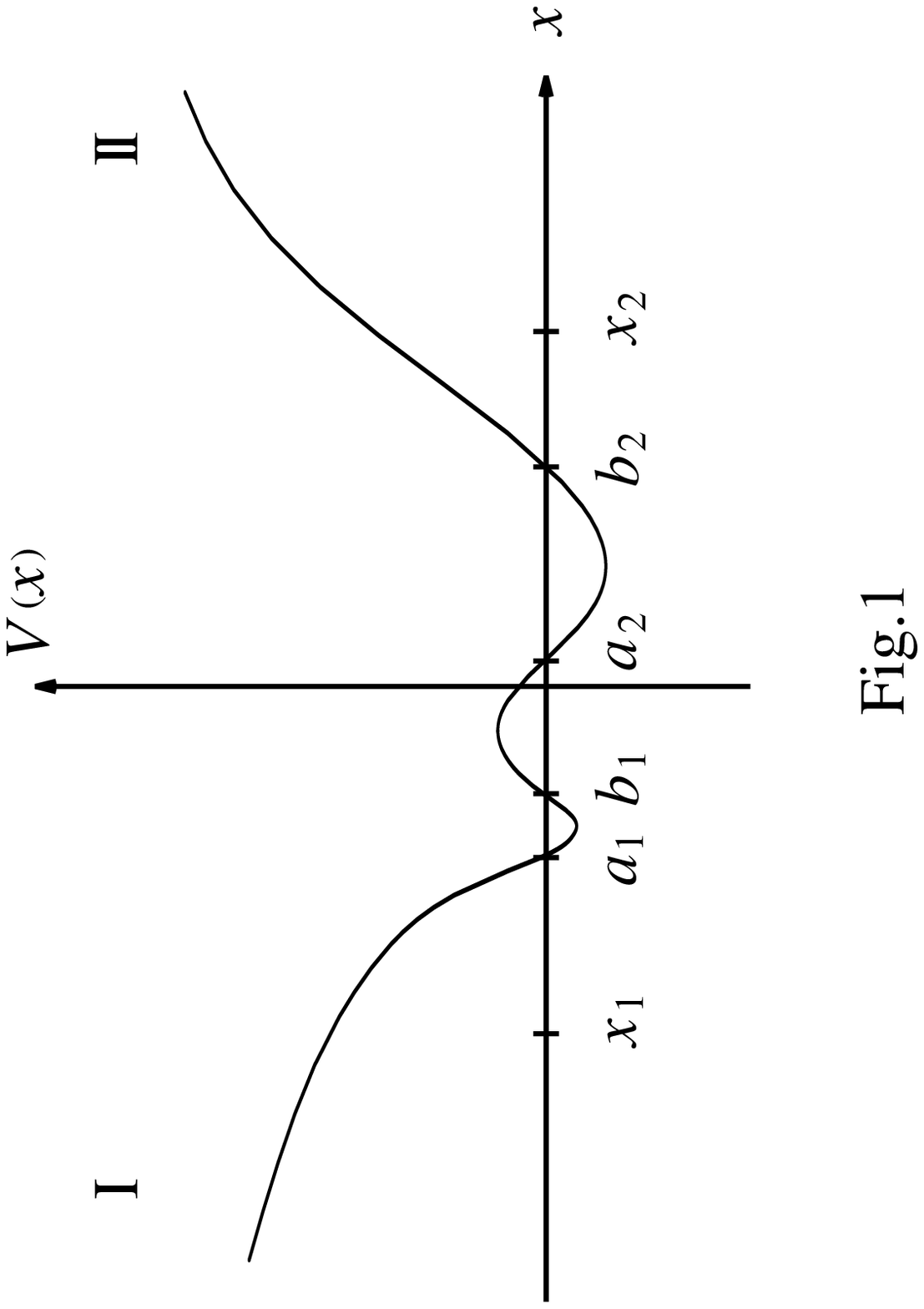}
}
\vfill
\eject

{}~~
\vfill
\centerline{
\epsfxsize=9.6cm
\epsfbox{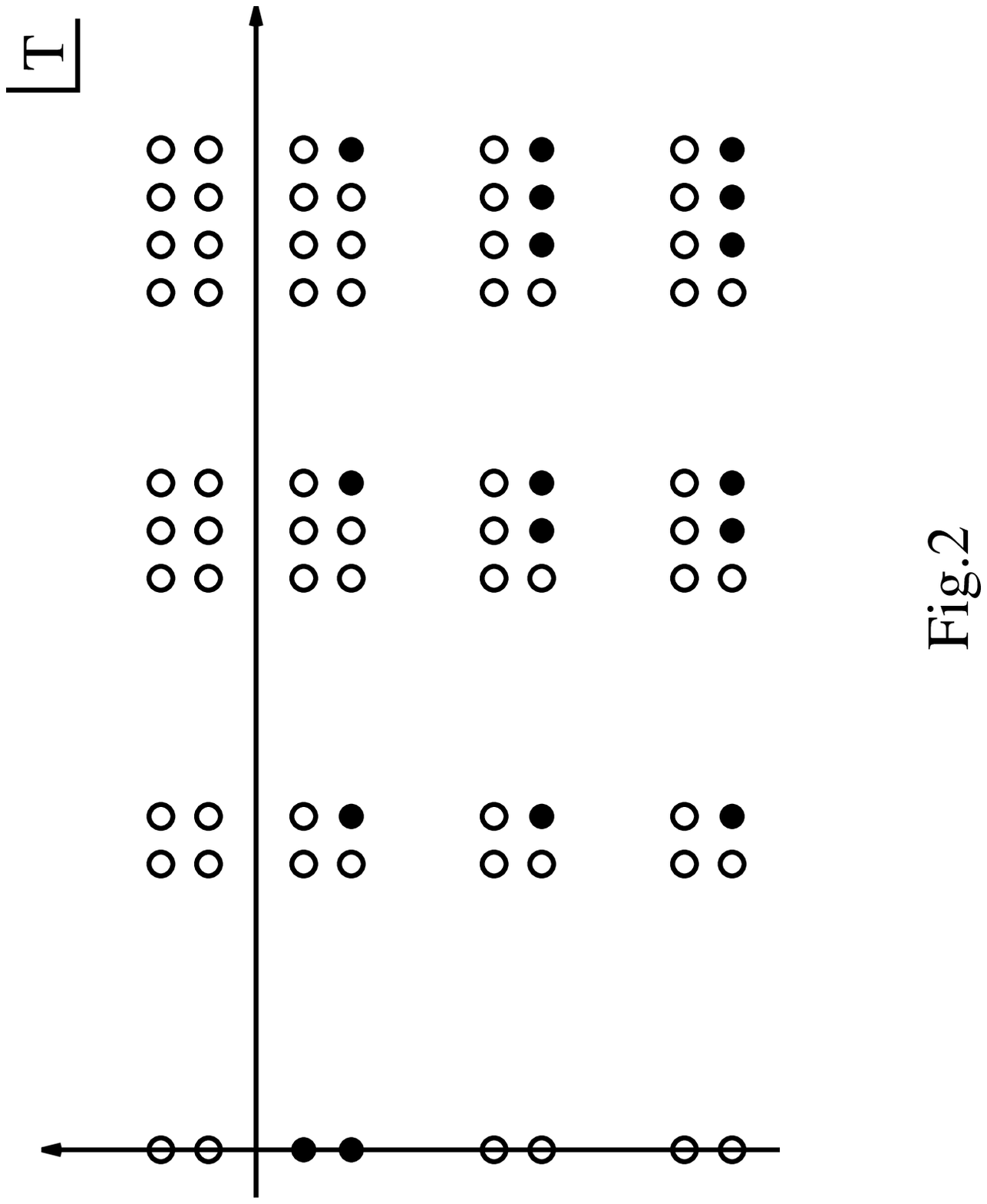}
}
\vfill
\eject

{}~~
\vfill
\centerline{
\epsfxsize=11.6cm
\epsfbox{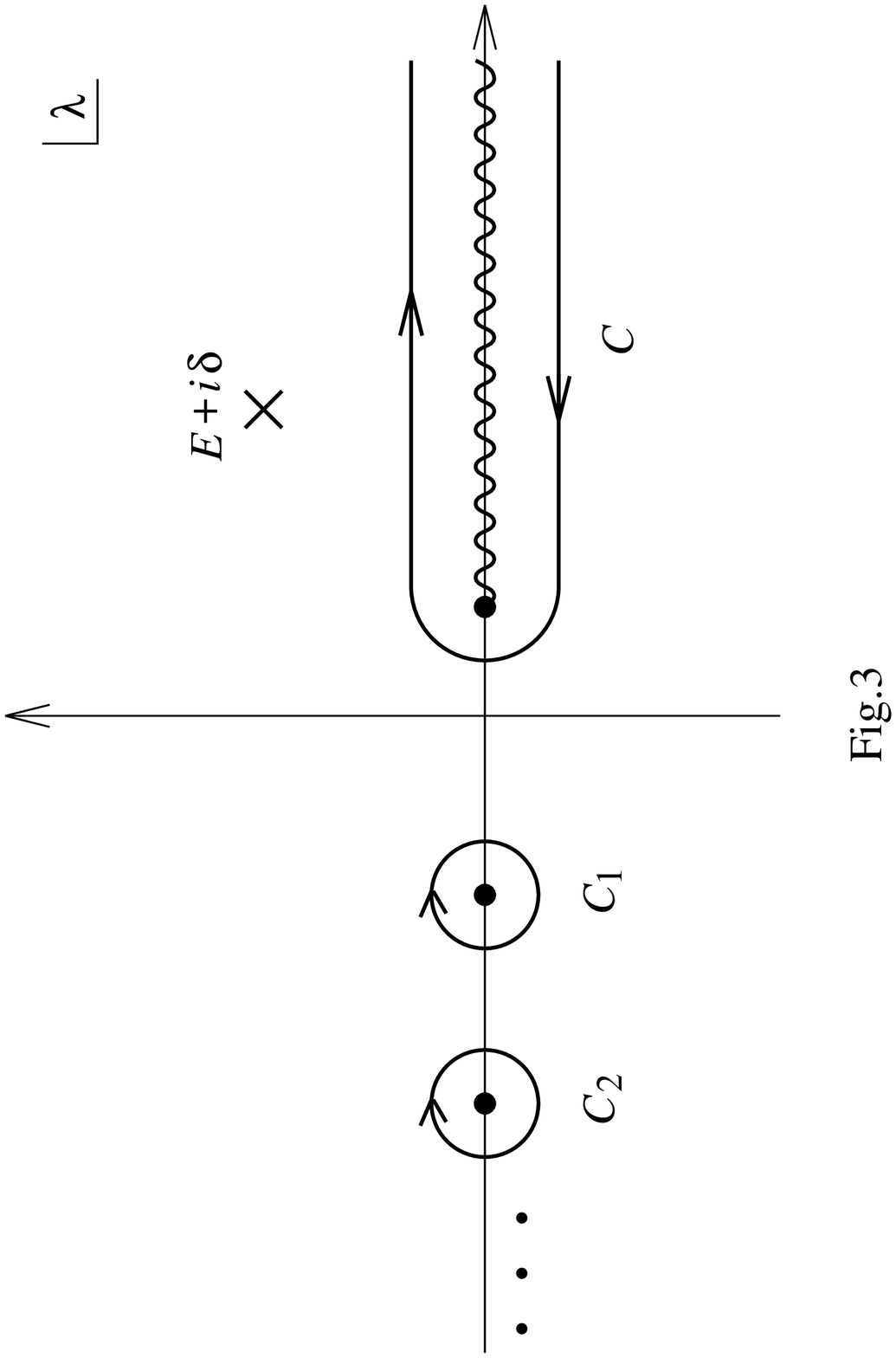}
}
\vfill\eject

{}~~
\vfill
\centerline{
\epsfxsize=9.6cm
\epsfbox{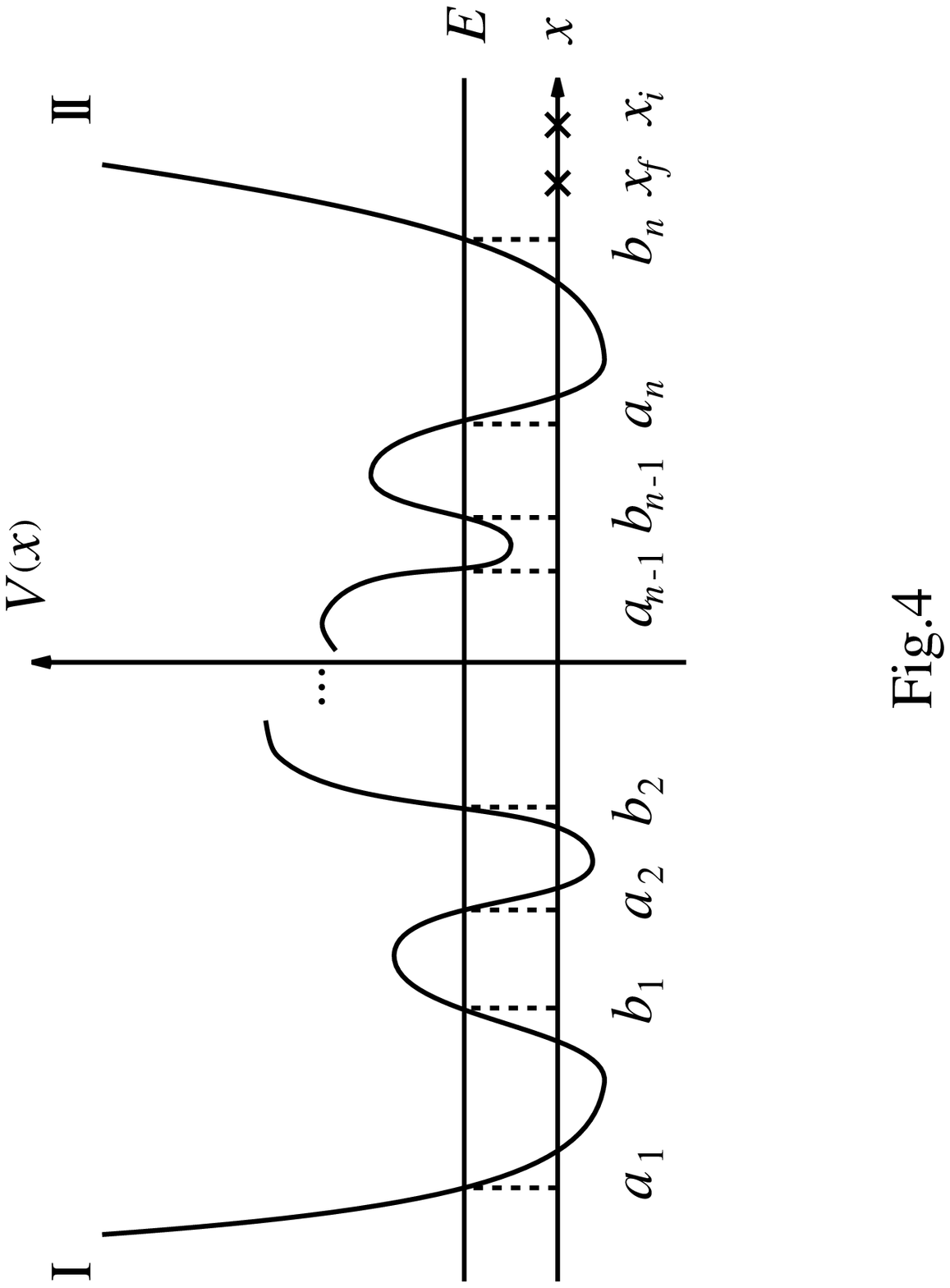}
}
\vfill\eject

{}~\vfill
\centerline{
\epsfxsize=10.8cm
\epsfbox{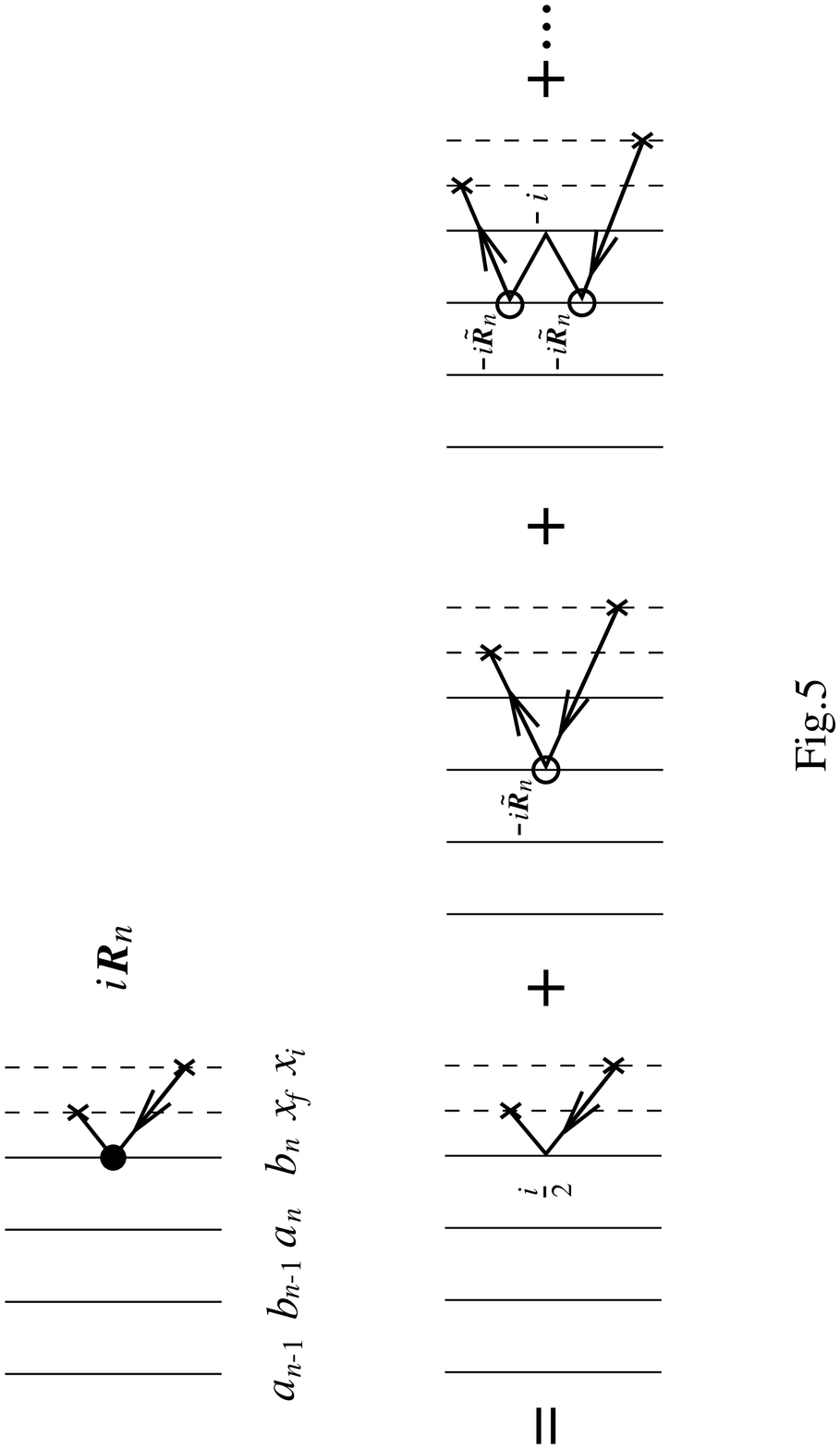}
}
\vfill\eject

{}~\vfill
\centerline{
\epsfxsize=10.8cm
\epsfbox{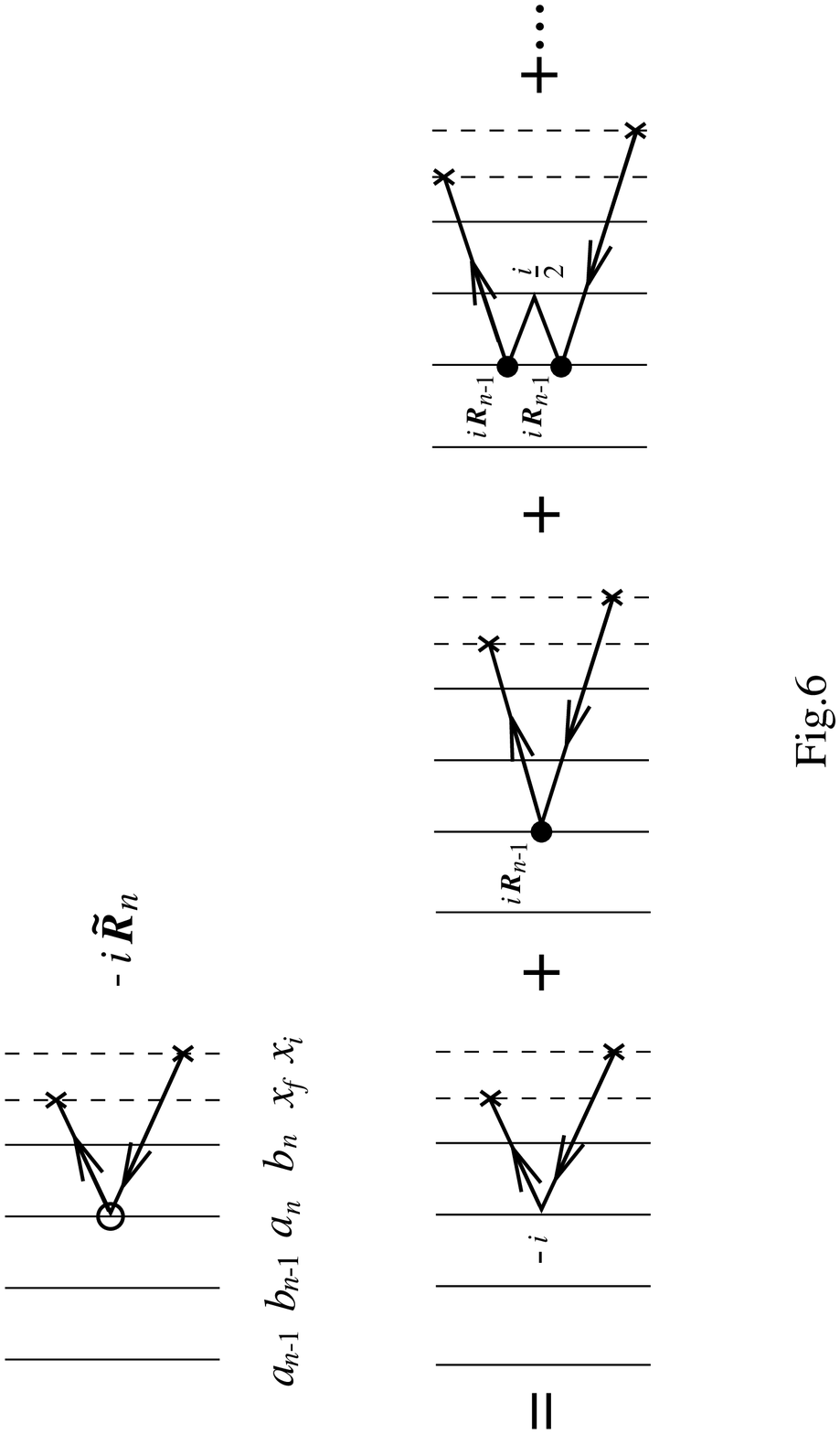}
}
\vfill\eject

{}~\vfill
\centerline{
\epsfxsize=11.7cm
\epsfbox{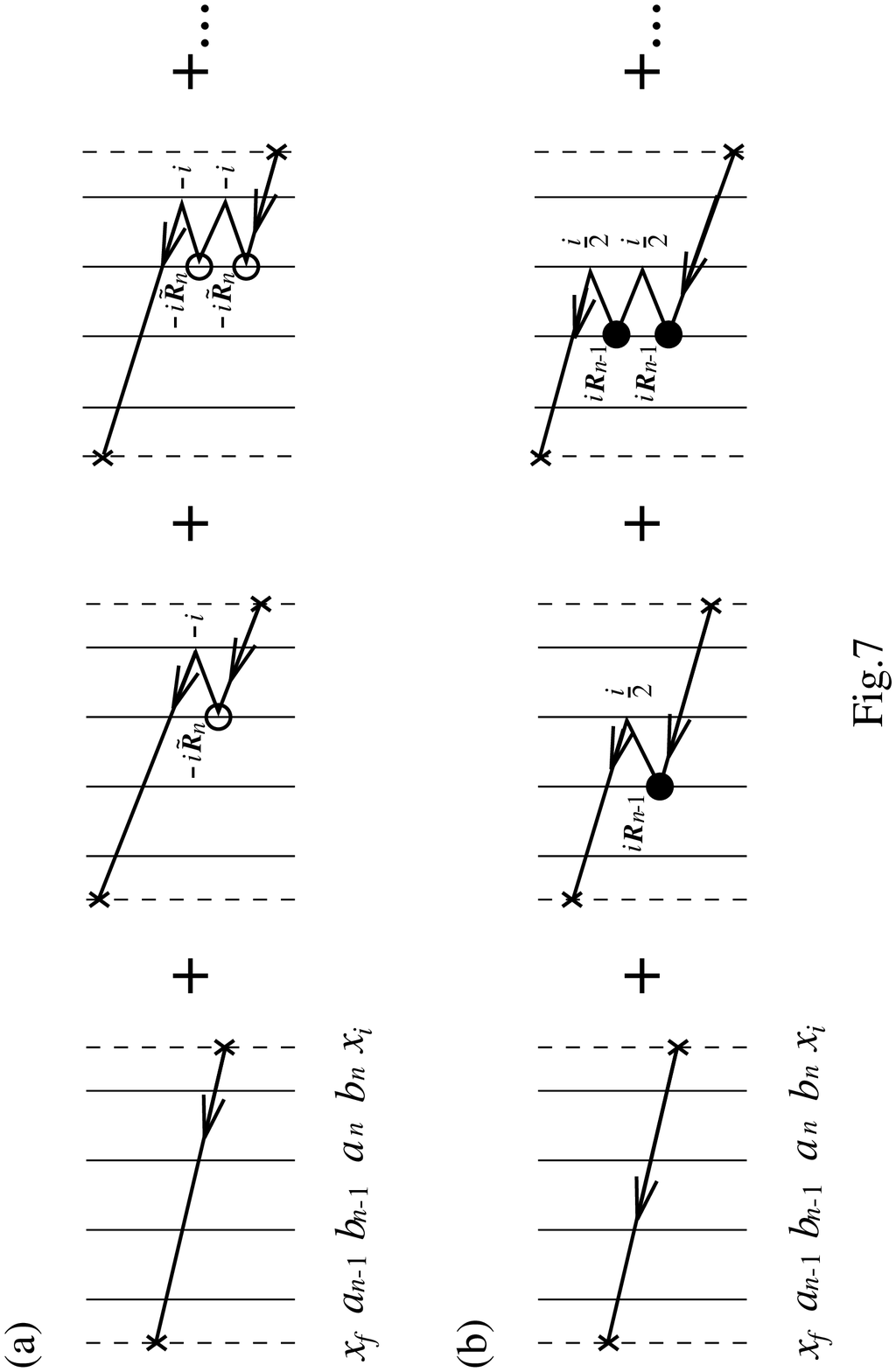}
}
\vfill\eject

{}~\vfill
\centerline{
\epsfxsize=11.3cm
\epsfbox{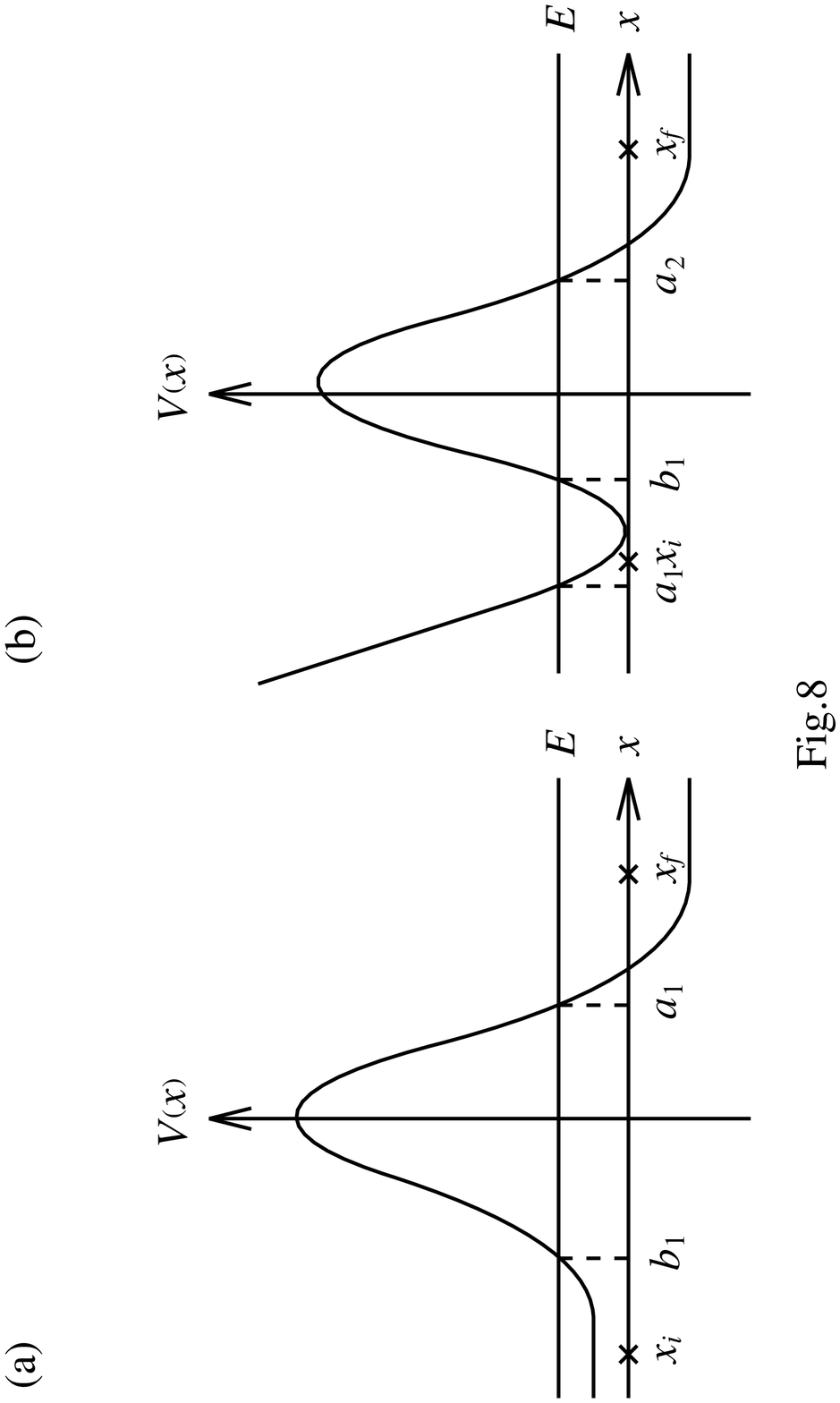}
}
\vfill\eject

\end